\documentclass[fleqn,twoside]{article}
\usepackage[headings]{espcrc2}
\usepackage{amsmath,amssymb,graphicx}

\newcommand\re{\mathrm{e}}
\newcommand\ri{\mathrm{i}}

\newcommand\eg{e.g.\ }
\newcommand\ie{i.e.\ }
\newcommand{\diag}{\mathop{\mathrm{diag}}}

\newcommand\bbC{\mathchoice
  {\setbox0=\hbox{$\displaystyle\mathrm{C}$}%
    \hbox{\hbox to 0pt{\kern0.4\wd0\vrule height .97\ht0\hss}\box0}}%
  {\setbox0=\hbox{$\textstyle\mathrm{C}$}%
    \hbox{\hbox to 0pt{\kern0.4\wd0\vrule height .97\ht0\hss}\box0}}%
  {\setbox0=\hbox{$\scriptstyle\mathrm{C}$}%
    \hbox{\hbox to 0pt{\kern0.4\wd0\vrule height .97\ht0\hss}\box0}}%
  {\setbox0=\hbox{$\scriptscriptstyle\mathrm{C}$}%
    \hbox{\hbox to 0pt{\kern0.4\wd0\vrule height .97\ht0\hss}\box0}}}
\newcommand\bbR{\mathrm{I\!R}}
\newcommand\Code[1]{\ensuremath{\texttt{%
  \def\_{\symbol{95}}%
  \def\{{\symbol{123}}%
  \def\}{\symbol{125}}%
  #1}}}

\newenvironment{itemlist}%
  {\begin{list}{$\bullet$}{\leftmargin=10bp}}%
  {\end{list}}

\allowdisplaybreaks

\title{Routines for the diagonalization of complex matrices}

\author{T. Hahn\address{%
	Max-Planck-Institut f\"ur Physik
\hfill MPP-2006-85 \\
	F\"ohringer Ring 6, D--80805 Munich, Germany
\hfill PACS: 02.10.Ud, 02.10.Yn, 02.60.Dc}}

\begin{document}

\begin{abstract}
Jacobi-type iterative algorithms for the eigenvalue decomposition,
singular value decomposition, and Takagi factorization of complex
matrices are presented.  They are implemented as compact Fortran 77
subroutines in a freely available library.
\end{abstract}

\maketitle


\section{Introduction}

This note describes a set of routines for the eigenvalue decomposition,
singular value decomposition, and Takagi factorization of a complex
matrix.  Unlike many other implementations, the current ones are all
based on the Jacobi algorithm, which makes the code very compact but
suitable only for small to medium-sized problems.

Although distributed as a library, the routines are self-contained and
can easily be taken out of the library and included in own code,
removing yet another installation prerequisite.  Owing to the small size
of the routines (each about 3 kBytes source code) it is possible, in
fact quite straightforward, to adapt the diagonalization routine to
one's own conventions rather than vice versa.


\section{Mathematical Background}

\subsection{Eigenvalue Decomposition}

The eigenvalue decomposition of a nonsingular matrix
$A\in\bbC^{n\times n}$ takes the form
\begin{equation}
U A\,U^{-1} = \diag(\sigma_1,\dots,\sigma_n)\,,\ 
\sigma_i\in\bbC\,.
\end{equation}
The eigenvalues $\sigma_i$ and transformation matrix $U$ can be further 
characterized if $A$ possesses certain properties:
\begin{itemize}
\item
$A = A^\dagger$ (Hermitian):
$U^{-1} = U^\dagger$,
$\sigma_i\in\bbR$,

\item
$A = A^T$ (symmetric):
$U^{-1} = U^T$.
\end{itemize}


\subsection{Singular Value Decomposition}

The singular value decomposition (SVD) can be applied to an arbitrary 
matrix $A\in\bbC^{m\times n}$:
\begin{gather}
V^* A\,W^\dagger = \diag(\sigma_1,\dots,\sigma_{\bar n})\,,
  \label{eq:svd} \\
V\in\bbC^{\bar n\times m},\quad
W^{-1} = W^\dagger\in\bbC^{n\times\bar n}, 
  \notag \\
\bar n = \min(m, n)\,,\quad \sigma_i\geqslant 0\,.
  \notag
\end{gather}
$V$ consists of orthonormal row vectors, \ie is also unitary for
$m = n$.


\subsection{Takagi Factorization}

The Takagi factorization \cite{Takagi,JohnsonHorn} is a less known
diagonalization method for complex symmetric matrices
$A = A^T\in\bbC^{n\times n}$,
\begin{gather}
U^* A\,U^\dagger = \diag(\sigma_1,\dots,\sigma_n)\,, \\
U^{-1} = U^\dagger\,,
\quad
\sigma_i \geqslant 0\,. \notag
\end{gather}
Although outwardly similar to the eigenvalue decomposition of a
Hermitian matrix, it is really the special case of an SVD with $V =
W^*$, as it applies even to singular matrices.  Note also that the left
and right factors, $U^*$ and $U^\dagger$, are in general not inverses of 
each other.

One might think that the Takagi factorization is merely a scaled SVD.  
For example, the matrix
\begin{equation}
A = \begin{pmatrix} 1 & 2 \\ 2 & 1 \end{pmatrix}
\end{equation}
has the SVD
\begin{gather}
V^T\diag(\sigma_1,\sigma_2) W = \\ \notag
\begin{pmatrix}
\tfrac 1{\sqrt 2} & \tfrac 1{\sqrt 2} \\[1ex]
\tfrac 1{\sqrt 2} & -\tfrac 1{\sqrt 2}
\end{pmatrix}^T
\begin{pmatrix}
3 & 0 \\
0 & 1
\end{pmatrix}
\begin{pmatrix}
\tfrac 1{\sqrt 2} & \tfrac 1{\sqrt 2} \\[1ex]
-\tfrac 1{\sqrt 2} & \tfrac 1{\sqrt 2}
\end{pmatrix}
\end{gather}
which can indeed be scaled to yield
\begin{gather}
U^T\diag(\sigma_1,\sigma_2) U = \\ \notag
\begin{pmatrix}
\tfrac 1{\sqrt 2} & \tfrac 1{\sqrt 2} \\[1ex]
\tfrac \ri{\sqrt 2} & -\tfrac \ri{\sqrt 2}
\end{pmatrix}^T
\begin{pmatrix}
3 & 0 \\
0 & 1
\end{pmatrix}
\begin{pmatrix}
\tfrac 1{\sqrt 2} & \tfrac 1{\sqrt 2} \\[1ex]
\tfrac \ri{\sqrt 2} & -\tfrac \ri{\sqrt 2}
\end{pmatrix}.
\end{gather}
But consider the matrix
\begin{equation}
A = \begin{pmatrix} 0 & 1 \\ 1 & 0 \end{pmatrix}
\end{equation}
which has the SVD
\begin{equation}
\begin{pmatrix} 1 & 0 \\ 0 & 1 \end{pmatrix}^T
\begin{pmatrix} 1 & 0 \\ 0 & 1 \end{pmatrix}
\begin{pmatrix} 0 & 1 \\ 1 & 0 \end{pmatrix}
\end{equation}
whereas its Takagi factorization is
\begin{equation}
\begin{pmatrix}
\tfrac 1{\sqrt 2} & \tfrac 1{\sqrt 2} \\[1ex]
-\tfrac \ri{\sqrt 2} & \tfrac \ri{\sqrt 2}
\end{pmatrix}^T
\begin{pmatrix}
1 & 0 \\
0 & 1
\end{pmatrix}
\begin{pmatrix}
\tfrac 1{\sqrt 2} & \tfrac 1{\sqrt 2} \\[1ex]
-\tfrac \ri{\sqrt 2} & \tfrac \ri{\sqrt 2}
\end{pmatrix}.
\end{equation}

Although occurring less frequently than the eigenvalue decomposition,
the Takagi factorization does have real applications in physics, \eg in
the diagonalization of mass matrices of Majorana fermions.


\section{Jacobi Algorithm}

The Jacobi algorithm \cite{Jacobi} consists of iteratively applying a
basic $2\times 2$ diagonalization formula until the entire $n\times n$
matrix is diagonal.  It works in several `sweeps' until convergence is
achieved.  In each sweep it rotates away the non-zero off-diagonal
elements using the $2\times 2$ algorithm.  Every such rotation of course
creates other non-zero off-diagonal elements.  It can be shown, however,
that the sum of the absolute values of the off-diagonal elements is
reduced in each sweep.  More precisely, the Jacobi method has quadratic
convergence \cite{Handbook}.

Convergence is in most cases achieved in $5{-}10$ sweeps, which for an
$n\times n$ matrix translates into $(10{-}20) n^3$ multiply--add
operations to obtain the eigenvalues only and $(15{-}30) n^3$ operations
including the eigenvectors (\cite{NR}, cf.\ also
Sect.~\ref{sect:timings}).  This compares with $\frac 23 n^3 + 30 n^2$
operations for the Householder/QL algorithm when just the eigenvalues
are sought and $\frac 43 n^3 + 3 n^3$ when also the eigenvectors are
needed.

For large $n$, the Jacobi algorithm is thus not the most efficient
method.  Nevertheless, for small to medium-sized problems the Jacobi
method is a strong competitor, in particular as it has the following
advantages:
\begin{itemlist}
\item It is conceptually very simple and thus very compact.

\item It delivers the eigenvectors at little extra cost.

\item The diagonal values are accurate to machine precision and,
      in cases where this is mathematically meaningful, the vectors 
      of the transformation matrix are always orthogonal, almost to 
      machine precision.
\end{itemlist}

For the various diagonalization problems discussed before, only the core 
$2\times 2$ diagonalization formula changes, whereas the surrounding
Jacobi algorithm stays essentially the same.

The famous Linear Algebra Handbook gives an explicit implementation of
the Jacobi algorithm for real symmetric matrices \cite{Handbook}, taking
particular care to minimize roundoff errors through mathematically
equivalent variants of the rotation formulas.  The present routines are
closely patterned on this procedure.  For the Takagi factorization, the
use of the Jacobi algorithm was first advocated in two conference papers
\cite{LathauwerMoor,WangQiao} which give only few details, however.  The
two-sided Jacobi version of the SVD is used in \cite{TwoSided}. 
Ref.~\cite{GolubVanLoan} gives a more thorough overview of literature
on the Jacobi method.


\section{The \boldmath{$2\times 2$} Formulas}

\subsection{Eigenvalue decomposition}

Using the ansatz
\begin{equation}
U = \begin{pmatrix}
c_1 & t_1 c_1 \\
-t_2 c_2 & c_2
\end{pmatrix}
\end{equation}
the equation $U A = \diag(\sigma_1, \sigma_2)\,U$ becomes
\begin{gather}
\sigma_1 = A_{11} + t_1 A_{21}
         = A_{22} + \frac 1{t_1} A_{12}\,, \label{eq:t1} \\
\sigma_2 = A_{11} - \frac 1{t_2} A_{21}    \label{eq:t2}
         = A_{22} - t_2 A_{12}\,.
\end{gather}
Solving for $t_1$ and $t_2$ yields
\begin{gather}
t_1 = \frac{A_{12}}{\Delta + D}\,,
\quad
t_2 = \frac{A_{21}}{\Delta + D}\,, \\
\Delta = \frac 12 (A_{11} - A_{22})\,, \\
D = \pm\sqrt{\Delta^2 + A_{12} A_{21}}\,.
\end{gather}
For the numerical stability it is best to choose the sign of $D$ 
which gives $t_{1,2}$ the larger denominator.  This corresponds to 
taking the smaller rotation angle ($< \pi/4$).  The diagonal values are
\begin{equation}
\begin{gathered}
\sigma_1 = A_{11} + \delta\,, \\
\sigma_2 = A_{22} - \delta\,,
\end{gathered}\qquad
\delta = \frac{A_{12} A_{21}}{\Delta + D}\,.
\end{equation}
In order that $U$ smoothly becomes unitary as $A$ becomes Hermitian, we 
choose
\begin{equation}
c_1 = c_2 = \frac 1{\sqrt{1 + t_1 t_2}}\,,
\end{equation}
which guarantees a unit determinant.


\subsection{Takagi Factorization}

Substituting the unitary ansatz
\begin{gather}
U = \begin{pmatrix}
c & t\,c\,\re^{\ri\varphi} \\
-t\,c\,\re^{-\ri\varphi} & c
\end{pmatrix}, \\
c = \frac 1{\sqrt{1 + t^2}}\,,\quad
t\in\bbR\,,
\end{gather}
into $U^* A = \diag(\sigma_1, \sigma_2)\,U$ and introducing
\begin{align}
\tilde\sigma_1 &= \re^{\ri\varphi}\sigma_1\,, &
\tilde A_{11} &= \re^{\ri\varphi} A_{11}\,, \\
\tilde\sigma_2 &= \re^{-\ri\varphi}\sigma_2\,, &
\tilde A_{22} &= \re^{-\ri\varphi} A_{22}\,,
\end{align}
we arrive at
\begin{gather}
\tilde\sigma_1 = \tilde A_{11} + t A_{12}
               = \tilde A_{22} + \frac 1t A_{12}\,, \\
\tilde\sigma_2 = \tilde A_{11} - \frac 1t A_{12}
               = \tilde A_{22} - t A_{12}\,.
\end{gather}
Comparing with Eqs.~\eqref{eq:t1} and \eqref{eq:t2}, the solution
can be read off easily:
\begin{gather}
t = \frac{A_{12}}{\tilde\Delta + \tilde D}\,, \\
\tilde\Delta = \frac 12 (\tilde A_{11} - \tilde A_{22})\,, \\
\tilde D = \pm\sqrt{\tilde\Delta^2 + A_{12}^2}\,.
\end{gather}
Again it is best for numerical stability to choose the sign of 
$\tilde D$ which gives the larger denominator for $t$.  The diagonal 
values become
\begin{gather}
\sigma_1 = A_{11} + t\,A_{12}\,\re^{-\ri\varphi}, \\
\sigma_2 = A_{22} - t\,A_{12}\,\re^{\ri\varphi}.
\end{gather}
The assumption $t\in\bbR$ fixes the phase $\varphi$.  It requires that
$A_{12}$ and $\tilde\Delta$ have the same phase, \ie $\tilde\Delta 
= \text{(real number)}\cdot A_{12}$.  Since both $\re^{\ri\varphi}$ and 
its conjugate appear in $\tilde\Delta$, we try the ansatz
\begin{equation}
\re^{\ri\varphi} = \alpha A_{12} + \beta A_{12}^*
\end{equation}
and choose coefficients to make the $A_{12}^*$ term in
\begin{align}
\tilde\Delta\propto\:
& (\alpha A_{11} - \beta^* A_{22}) A_{12} + \notag \\
& (\beta A_{11} - \alpha^* A_{22}) A_{12}^*
\end{align}
vanish.  This is achieved by $\alpha = A_{11}^*$ and $\beta = A_{22}$ 
which also makes the coefficient of $A_{12}$ real.  Thus,
\begin{equation}
\re^{\ri\varphi} = \frac{A_{11}^* A_{12} + A_{22} A_{12}^*}
  {|A_{11}^* A_{12} + A_{22} A_{12}^*|}\,.
\end{equation}


\section{Singular Value Decomposition}

We insert unitary parameterizations for the left and right
transformation matrices $X = V,W$,
\begin{gather}
X = \begin{pmatrix}
c_X & t_X\,c_X \\
-t_X^*\,c_X & c_X
\end{pmatrix}, \\
c_X = \frac 1{\sqrt{1 + |t_X|^2}}\,, \quad
t_X\in\bbC\,,
\end{gather}
into $V^* A = \diag(\sigma_1, \sigma_2)\,W$ and by eliminating 
$\sigma_{1,2}$ arrive at
\begin{align}
A_{12} + A_{22} t_V^* &= (A_{11} + A_{21} t_V^*) t_W\,, \\
A_{21} + A_{22} t_W^* &= (A_{11} + A_{12} t_W^*) t_V\,.
\end{align}
The solutions are evidently related through exchange of the off-diagonal 
elements.  Explicitly,
\begin{gather}
t_V = \frac{M_V}{\Delta_V + D_V}\,, \\
M_V = A_{11}^* A_{21} + A_{22} A_{12}^*\,, \\
\Delta_V = \frac 12 (|A_{11}|^2 - |A_{22}|^2 + \notag \\[-1ex]
\hspace*{4em}        |A_{12}|^2 - |A_{21}|^2)\,, \\
D_V = \pm\sqrt{\Delta_V^2 + |M_V|^2}\,, \\
t_W = \left.t_V\right|_{A_{12}\leftrightarrow A_{21}}.
\end{gather}
$D_V$ and $D_W$ share the same sign, which is chosen to yield the
larger set of denominators for better numerical stability.

The singular values become
\begin{gather}
\sigma_1 = \frac{c_V}{c_W} (A_{11} + t_V^* A_{21})\,, \\
\sigma_2 = \frac{c_V}{c_W} (A_{22} - t_V A_{12})\,, \label{eq:sig2}
\end{gather}

If $A\in\bbC^{m\times n}$ is not a square matrix, we consider two cases:

For $m > n$, we make $A$ square by padding it with zero-columns at the
right.  For zero right column, $A_{12} = A_{22} = 0$,
Eq.~(\ref{eq:sig2}) guarantees that it is $\sigma_2$ that vanishes. 
That is, the above Jacobi rotation never moves a singular value into the
zero-extended part of the matrix.  All singular values automatically end
up as the \emph{first} $\min(m,n)$ diagonal values of the Jacobi-rotated
$A$.

For $m < n$, we apply this algorithm to $A^T$ and at the end exchange
$V$ and $W$.  This is the least involved solution, as $A$ has to be
copied to temporary storage for zero-extension anyway.


\section{Installation}

The Diag package can be downloaded from the URL
\Code{http://www.feynarts.de/diag}.  After unpacking the tar file, the
library is built with
\begin{verbatim}
  ./configure
  make
\end{verbatim}
To compile also the Mathematica executable, one needs to issue
``\Code{make all}'' instead of just ``\Code{make}.''  The generated
files are installed into a platform-dependent directory with
``\Code{make install}'' and at the end one can do a ``\Code{make
clean}'' to remove intermediate files.

The routines in the Diag library allocate space for intermediate results
according to a preprocessor variable \Code{MAXDIM}, defined in
\Code{diag.h}.  This effectively limits the size of the input and output
matrices but is necessary because Fortran 77 offers no dynamic memory
allocation.  Since the Jacobi algorithm is not particularly suited for
large problems anyway, the default value of 16 should be sufficient for
most purposes.


\section{Description of the Fortran Routines}

The general convention is that each matrix is followed by its leading
dimension in the argument list, \ie the $m$ in \Code{A($m$,$n$)}.  In
this way it is possible to diagonalize submatrices with just a different
invocation.  Needless to add, the leading dimension must be at least as
large as the corresponding matrix dimension.


\subsection{Hermitian Eigenvalue Decomposition}

Hermitian matrices are diagonalized with
\begin{verbatim}
subroutine HEigensystem(n, A,ldA,
                        d, U,ldU, sort)
integer n, ldA, ldU, sort
double complex A(ldA,n), U(ldU,n)
double precision d(n)
\end{verbatim}
The arguments are as follows:
\begin{itemlist}
\item
\Code{n} (input), the matrix dimension.

\item
\Code{A} (input), the matrix to be diagonalized.  Only the upper
triangle of \Code{A} needs to be filled and it is further assumed that
the diagonal elements are real.  Attention: the contents of \Code{A} are 
not preserved.

\item
\Code{d} (output), the eigenvalues.

\item
\Code{U} (output), the transformation matrix.

\item
\Code{sort} (input), a flag that determines sorting of the eigenvalues:
\begin{align*}
 0 &= \text{do not sort}, \\
 1 &= \text{sort into ascending order}, \\
-1 &= \text{sort into descending order}.
\end{align*}
The `natural' (unsorted) order is determined by the choice of the
smaller rotation angle in each Jacobi rotation.
\end{itemlist}


\subsection{Symmetric Eigenvalue Decomposition}

The second special case is that of a complex symmetric matrix:
\begin{verbatim}
subroutine SEigensystem(n, A,ldA,
                        d, U,ldU, sort)
integer n, ldA, ldU, sort
double complex A(ldA,n), U(ldU,n)
double complex d(n)
\end{verbatim}
The arguments have the same meaning as for \Code{HEigensystem}, except 
that \Code{A}'s diagonal elements are not assumed real and sorting 
occurs with respect to the real part only.


\subsection{General Eigenvalue Decomposition}

The general case of the eigenvalue decomposition is implemented in
\begin{verbatim}
subroutine CEigensystem(n, A,ldA,
                        d, U,ldU, sort)
integer n, ldA, ldU, sort
double complex A(ldA,n), U(ldU,n)
double complex d(n)
\end{verbatim}
The arguments are as before, except that \Code{A} has to be filled 
completely.


\subsection{Takagi Factorization}

The Takagi factorization is invoked in almost the same way as
\Code{SEigensystem}:
\begin{verbatim}
subroutine TakagiFactor(n, A,ldA,
                        d, U,ldU, sort)
integer n, ldA, ldU, sort
double complex A(ldA,n), U(ldU,n)
double precision d(n)
\end{verbatim}
The arguments are as for \Code{SEigensystem}.  Also here only the 
upper triangle of \Code{A} needs to be filled.


\subsection{Singular Value Decomposition}

The SVD routine has the form
\begin{verbatim}
subroutine SVD(m, n, A,ldA,
               d, V,ldV, W,ldW, sort)
integer m, n, ldA, ldV, ldW, sort
double complex A(ldA,n)
double complex V(ldV,m), W(ldW,n)
double precision d(min(m,n))
\end{verbatim}
with the arguments
\begin{itemlist}
\item \Code{m}, \Code{n} (input), the dimensions of \Code{A}.
\item \Code{A} (input), the $\Code{m}\times\Code{n}$ matrix of which the 
      SVD is sought.
\item \Code{d} (output), the singular values.
\item \Code{V} (output), the $\min(\Code{m},\Code{n})\times\Code{m}$
      left transformation matrix.
\item \Code{W} (output), the $\min(\Code{m},\Code{n})\times\Code{n}$
      right transformation matrix.
\item \Code{sort} (input), the sorting flag with values as above.
\end{itemlist}


\section{Description of the C Routines}

The C version consists merely of an include file \Code{CDiag.h}
which sets up the correct interfacing code for using the Fortran
routines.  In particular the usual problem of transposition\footnote{%
	C uses row-major storage for matrices, whereas Fortran uses 
	column-major storage, \ie a matrix is a vector of row vectors 
	in C, and a vector of column vectors in Fortran.}
between Fortran and C is taken care of.

The arguments are otherwise as for Fortran.  To treat complex numbers
uniformly in C and C++, \Code{CDiag.h} introduces the new
\Code{double\_complex} type which is equivalent to
\Code{complex<double>} in C++ and to \Code{struct \{ double re, im; \}}
in C.

C's syntax unfortunately does not allow the declaration of variable-size
matrices as function arguments, thus it is not possible for
\Code{CDiag.h} to set up the prototypes to directly accept C matrices
without compiler warnings.  To simplify matters, \Code{CDiag.h} defines
the macro \Code{Matrix} which is used as in
\begin{verbatim}
double_complex A[5][3], V[3][5], W[3][3];
double d[3];
...
SVD(5, 3, Matrix(A), d,
          Matrix(V), Matrix(W), 0);
\end{verbatim}
This is equivalent to using an explicit cast and passing the leading
dimension, \ie
\begin{verbatim}
SVD(5, 3, (double_complex *)A,3, d,
          (double_complex *)V,5,
          (double_complex *)W,3, 0);
\end{verbatim}

Compilation should be done using the included \Code{fcc} script, \ie by 
replacing the C compiler with \Code{fcc}, for example
\begin{verbatim}
fcc -Iinclude myprogram.c -Llib -ldiag
\end{verbatim}
The \Code{fcc} script is configured and installed during the build 
process and automatically adds the necessary flags for linking with 
Fortran code.


\section{The Mathematica Interface}

The Mathematica version may not seem as useful as the Fortran library
since Mathematica already has perfectly functional eigen- and singular
value decompositions.  The Takagi factorization is not available in
Mathematica, however, and moreover the interface is ideal for trying
out, interactively using, and testing the Diag routines.

The Mathematica executable is loaded with
\begin{verbatim}
  Install["Diag"]
\end{verbatim}
and makes the following functions available:


\begin{itemlist}
\item
\Code{HEigensystem[A]} computes the eigenvalue decomposition
\Code{\{d,\,U\}} of the Hermitian matrix \Code{A} such that
\Code{U.A == DiagonalMatrix[d].U}.

\item
\Code{SEigensystem[A]} computes the eigenvalue decomposition
\Code{\{d,\,U\}} of the symmetric matrix \Code{A} such that
\Code{U.A == DiagonalMatrix[d].U}.

\item
\Code{CEigensystem[A]} computes the eigenvalue decomposition
\Code{\{d,\,U\}} of the general matrix \Code{A} such that
\Code{U.A == DiagonalMatrix[d].U}.

\item
\Code{TakagiFactor[A]} computes the Takagi factorization
\Code{\{d,\,U\}} of the symmetric matrix \Code{A} such that \\
\Code{Conjugate[U].A == DiagonalMatrix[d].U}.

\item
\Code{SVD[A]} computes the singular value decomposition
\Code{\{V,\,d,\,W\}} of the matrix \Code{A} such that
\Code{Conjugate[V].A == DiagonalMatrix[d].W}.
\end{itemlist}
Note that these routines do not check whether the given matrix fulfills
the requirements, \eg whether it is indeed Hermitian.


\section{Timings}
\label{sect:timings}

The following plot shows the time for diagonalizing a random matrix of
various dimensions.  Note that the abscissa is divided in units of the
dimension cubed; this accounts for the anticipated scaling behaviour 
of the Jacobi algorithm, hence the curves appear essentially linear.

The absolute time values should be taken for orientation only, as they
necessarily reflect the CPU speed.  For reference, the numbers used in
the figure above were obtained on an AMD64-X2 CPU running at 3 GHz. 
Each point is actually the average from diagonalizing $10^6$ random
matrices, to reduce quantization effects in the time measurement.

\begin{center}
\includegraphics[width=\linewidth]{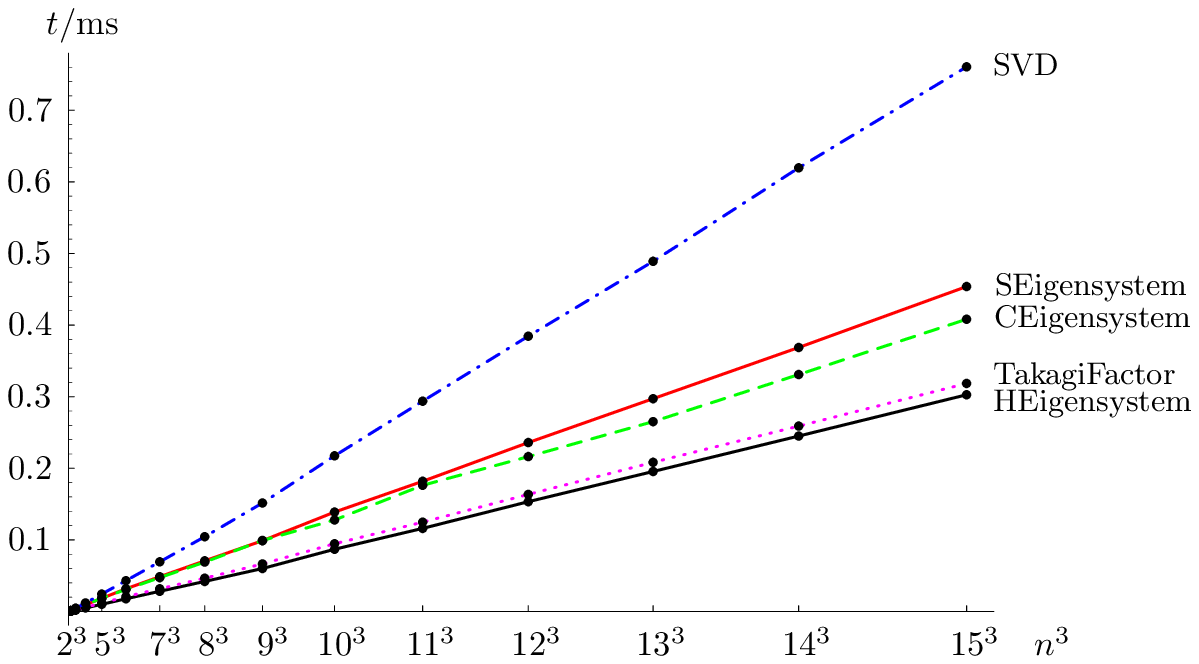}
\end{center}

The average number of sweeps needed to diagonalize the $10^6$ random 
matrices to machine precision is plotted in the next figure.

\begin{center}
\includegraphics[width=\linewidth]{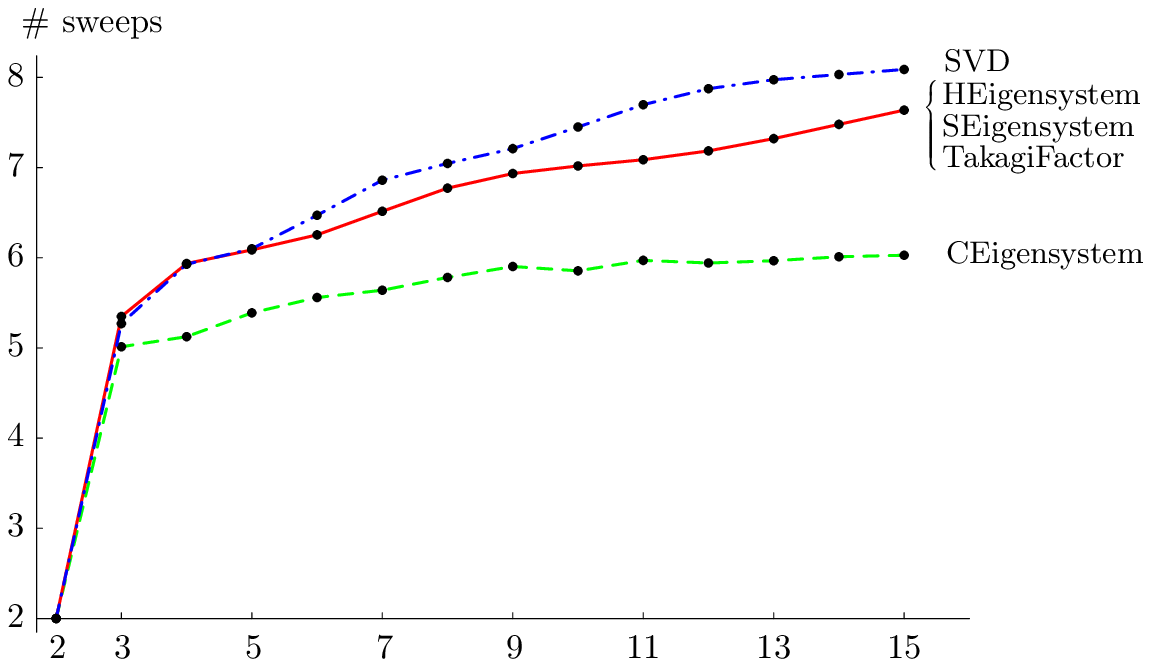}
\end{center}


\section{Summary}

The Diag package contains Fortran subroutines for the eigenvalue 
decomposition, singular value decomposition, and Takagi factorization of 
a complex matrix.  The Fortran library is supplemented by interfacing 
code to access the routines from C/C++ and Mathematica.

The routines are based on the Jacobi algorithm.  They are self-contained
and quite compact, thus it should be straightforward to use them outside
of the library.  All routines are licensed under the LGPL.


\section*{Acknowledgements}

TH thanks the National Center for Theoretical Studies, Hsinchu, Taiwan,
for warm hospitality during the time this work was carried out.


\raggedright

\end{document}